# Probing the phonon surface interaction by wave packet simulation: effect of roughness and morphology

Cheng Shao[1], Qingyuan Rong[1], Ming Hu[2,3], Hua Bao[1,*]

1 University of Michigan—Shanghai Jiao Tong University Joint Institute, Shanghai Jiao Tong University, Shanghai

2 Institute of Mineral Engineering, Division of Materials Science and Engineering, Faculty of Georesources and Materials Engineering, RWTH Aachen University, Aachen 52064, Germany

3Aachen Institute for Advanced Study in Computational Engineering Science (AICES), RWTH Aachen University, Aachen 52062, Germany

* Corresponding author: hua.bao@sjtu.edu.cn

# Abstract

One way to reduce the lattice thermal conductivity of solids is to induce additional phonon surface scattering through nanostructures. However, how phonons interact with boundaries, especially at the atomic level, is not well understood. In this work, we performed two-dimensional atomistic wave packet simulations to investigate the phonon surface interaction. Emphasis has been given to the angular-resolved phonon reflection at smooth, periodically rough, and amorphous surfaces. We found that the acoustic phonon reflection at a smooth surface is not simply specular. Mode conversion can occur after reflection, and the detailed energy distribution after reflection will dependent on surface condition and polarization of incident phonon. At periodically rough surfaces, the reflected wave packet distribution does not follow the well-known Ziman's model, but shows a nonmonotonic dependence on the depth of surface roughness. When an amorphous layer is attached to the surface, the incident wave packet will be absorbed by the amorphous region, and results in quite diffusive reflection. Our results clearly show that the commonly used specular-diffusive model is not enough to describe the phonon reflection at a periodically rough surface, while an amorphous layer can induce strong diffusive reflection. This work provides a careful analysis of phonon reflection at a surface with different morphology, which is important to a better understanding of thermal transport in various nanostructures.

# I. Introduction

Engineering the thermal conductivity of solids is an important research topic in the past two decades. Driven by the requirement of more efficient thermoelectric materials (i.e., larger figure of merit, or *ZT*), low lattice thermal conductivity is highly desirable. A fundamental idea is to utilize phonon surface scattering in nanostructures to reduce the lattice thermal conductivity. Tremendous efforts have been devoted into the theoretical and experimental investigations of thermal transport in various types of nanostructures, including embedded nanoparticles[1], nanowires[2–5], thin films[6–9] and porous structures[10–13]. Significant breakthrough in reducing thermal conductivity has been achieved. For example, the thermal conductivity of crystalline $In_{0.53}Ga_{0.47}As$ can be lower than the "alloy limit" by embedding ErAs nanoparticle[1]. The thermal conductivities of rough Si nanowires are found to be two orders of magnitude lower than the bulk counterpart[3]. Thermal conductivities of silicon thin films with periodically arranged pores are also reported to be two orders of magnitude lower than that of bulk silicon[10–12].

Despite the numerous research efforts, a fundamental question remains unclear, i.e., how exactly are phonons scattered at surfaces in nanostructures? Thermal conductivity reduction in nanostructures are mostly attributed to the classical size effect: surfaces can induce additional scattering channels, which limit the phonon mean free path and thus reduce the lattice thermal conductivity[14]. At a surface, phonons can be reflected and transmitted. In comparison to the numerous research on phonon transmission in atomistic systems[15–18], phonon reflection in atomistic systems is less explored[19]. It is usually assumed that after scattering at surface, the phonon reflection shows a specular component and a diffuse component[14,20]. The portion of the energy that is reflected in the specular direction is represented by the specularity parameter $p$, while the other part (1- $p$) of energy is diffusely scattered.

In the 1955, Ziman proposed an analytical formula (known as Ziman's formula) to estimate the specularity parameters at surfaces[21,20], which is later extended by Soffer[22] to include angular dependence. Since then this formula has been applied to describe the phonon scattering at surfaces[23–26]. Recent studies showed that this formula is only valid under the Kirchhoff approximation[27,28]. In some other investigations, the specularity parameter is simply taken as a fitting parameter to match the theory and experiments[29–33]. On the other hand, based on Boltzmann transport theory, the lowest thermal conductivity of a nanowire should occur when the specularity parameter is 0 (i.e., completely diffuse scattering), which is also known as the "Casimir limit"[34]. However, recent measurements show that the thermal conductivity of rough nanowires is even lower than the "Casimir limit"[3]. These new experimental evidences motivate many recently attempts to develop a new theory beyond the traditional understanding[4,5,35–40], especially on how surface roughness can affect the phonon scattering and reduce the lattice thermal conductivity of nanowires by two orders of magnitude. Most of these researches were based on continuum models: some assumed the scattering process is particle-like[5,35] while others considered the wave nature of phonon[36,41,42]. Although these works significantly advance our understanding of phonon surface scattering, how the atomic-level surface morphology can affect phonon surface scattering is still an open question.

In this work, we investigated angular-resolved phonon reflection at surfaces using the wave packet simulation. This method uses molecular dynamics to study the phonon surface scattering with an atomistic approach and requires no assumption other than the interatomic potential[15,43–46]. Instead of the widely used quasi-one-dimensional simulation, a two-dimensional wave packet with oblique incidence[19] is built to investigate the phonon scattering at different types of surfaces. Three types of surfaces have been considered in this work: 1) fixed or free smooth surface; 2) surface with different periodic roughness and 3) surface with a layer of amorphous structure. The atomic level details of the phonon scattering process can be extracted and visualized. The obtained results are compared with existing theories.

# II. Methods and simulation details

## A. Wave packet method

Molecular dynamics (MD) based wave packet method is applied to investigate the phonon surface scattering process. The wave packet method was first introduced by Schelling et al.[15] to study the phonon scattering at interfaces and subsequently applied to study a wide area of problems[16,44,45]. This method requires no prior knowledge of phonon surface scattering.

A wave packet is a linear combination of plane waves with certain polarization and centered at a specified wavevector. To generate a phonon wave packet centered at position $\mathbf{x}_0$ with a wave vector $\mathbf{k}$ and polarization $v$, the atomic displacement $\mathbf{u}_{l,b}$ and velocity $\mathbf{v}_{l,b}$ of $b^{th}$ atom in the $l^{th}$ unit cell can be expressed as following [15,16]:

$$\mathbf{u}_{l,b} = \frac{A}{\sqrt{m_b}} \boldsymbol{\varepsilon}_{v,\mathbf{k},b} \exp[i(\mathbf{k}\cdot(\mathbf{x}_l - \mathbf{x}_0) - \omega t)] \exp[-\frac{(\mathbf{x}_l - \mathbf{x}_0 - \mathbf{v}_g t)^2}{\sigma^2}], \quad (1)$$

$$\mathbf{v}_{l,b} = \frac{\partial \mathbf{u}_{l,b}}{\partial t} = (-i\omega + \frac{2(\mathbf{x}_l - \mathbf{x}_0)\cdot \mathbf{v}_g}{\sigma^2})\mathbf{u}_{l,b}, \quad (2)$$

where $A$ is the amplitude of the wave packet, $m_b$ is the atomic mass of $b^{th}$ atom in a unit cell, and $\boldsymbol{\varepsilon}_{v,\mathbf{k},b}$ is the eigenvector for the $b^{th}$ atom in a unit cell with wave vector $\mathbf{k}$ and polarization $v$, which can be obtained by lattice dynamics calculation[47] using the GULP package[48]. $\mathbf{x}_l$ is the position of the $l^{th}$ unit cell, $\mathbf{x}_0$ is the center position of the wave packet, $\omega$ is the angular frequency of the phonon mode, $\mathbf{v}_g$ is the group velocity of this mode, and $\sigma$ is the spatial extent of the wave packet. The initial atomic velocity of the wave packet is the time derivative of the initial displacement. One should note that the initial displacements and velocities calculated from Eq. (1) and Eq. (2) are complex numbers. In our simulations, the real part of the displacements and velocities are used to generate the wave packet.

To obtain a deeper insight about the dynamics of wave packet after scattering, we also projected the real space atomic vibrations into reciprocal space and showed how the reflected wave energy distributed in reciprocal space. The associated kinetic energy at wavevector $\mathbf{k}$ and time $t$ in reciprocal space can be represented as [49–51]:

$$\Phi(\mathbf{k},t) = \sum_{\alpha,b}^{3,n} \frac{m_b}{N} \left| \sum_l^N \dot{u}_{l,b}^{\alpha}(t) \exp(i\mathbf{k}\cdot \mathbf{r}_0^l) \right|^2, \quad (3)$$

where $\Phi(\mathbf{k},t)$ is the kinetic energy in reciprocal space at wavevector $\mathbf{k}$ and time $t$, $m_b$ is the mass of $b^{th}$ atom in the unit cell, $N$ is number of unit cells in the simulation domain, $\dot{u}_{l,b}^{\alpha}(t)$ is the

$\alpha$ component of the velocity of $b^{\text{th}}$ atom in the $l^{\text{th}}$ unit cell at time $t$. The first summation is over different atom $b$ and Cartesian coordinate directions $\alpha$ in a unit cell. The second summation is over all the unit cells $l$ in a system. To obtain the energy distribution in reciprocal space, we discretized the first Brillouin zone with a $k$ increment of $0.02\times 2\pi/a$.

## B. Simulation details

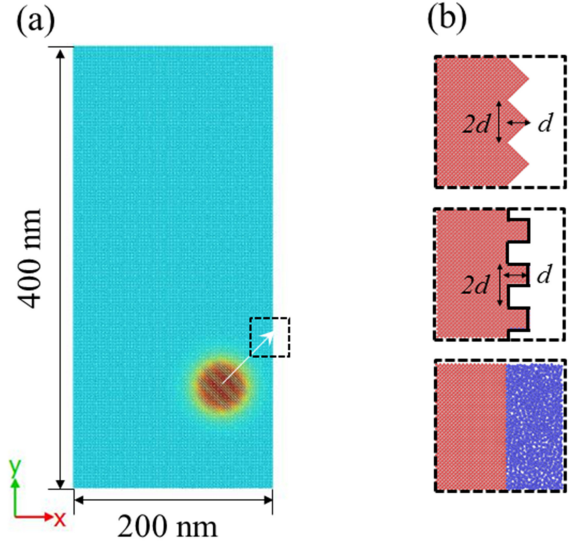

FIG. 1. (a) Schematic diagram of the simulation domain. (b) Atomic structures of different types of surfaces and boundary conditions. From top to bottom are sawtooth shape surface with free boundary condition, square shape surface with fixed boundary condition, and smooth surface adjacent by an amorphous layer. $d$ represent the depth of the surface roughness and $2d$ is the period length.

Silicon is selected as our model system due to numerous studies on silicon-based nanostructure[2,10,26,52]. The MD simulations were performed in LAMMPS [53] using Tersoff potential [54] with a time step of 1 fs. To model oblique incidence, two-dimensional wave packet needs to be generated, requiring a large in-plane simulation domain. We used a simulation domain with a dimension of $200\times 400$ nm$^2$ in the $xy$ plane (shown in Fig. 1(a)), in which the $x$ axis is aligned with the [100] crystal direction. To reduce the computational cost, the thickness of the simulation domain is chosen to be 4 unit cells (about 2 nm) with periodic boundary condition. Three types of surfaces are considered in this work, namely smooth surface, periodically rough surface, and a smooth surface adjacent by an amorphous layer. For each type of surface, we consider two different boundary conditions: free and fixed. For free boundary, the atoms at the surface are free to move while for fixed boundary condition the atoms in the outermost layer of surface are fixed. For periodically rough surface, two types of roughness are considered: the square shape and sawtooth shape. The rough surface is characterized by the period length $2d$ and depth $d$, as shown in Fig. 1(b). In our simulations, we are varying $d$ in the range of 5 to 40 Å to model different roughness. The thickness of the amorphous layer is about 10 nm. To generate this amorphous layer, we melted the atoms in this region at 6000K and annealed at rate of $10^{12}$ K/s while keeping the other atoms frozen, similar to our previous work[17]. After that we fully relaxed the whole system. The initial displacements and velocities of each atom are then calculated from Eq. (1) and Eq. (2)

for a specified wave packet and then added to the relaxed structure. The system then runs in an NVE ensemble and the kinetic and potential energy of each atom are monitored and dumped for post-process and analysis.

A localized wave packet is a superposition of waves with slightly different wavevectors. In a dispersive medium, waves with different wavevectors have slightly different group velocities $\mathbf{v}_g$, so it is expected that the wave packet will disperse as it travels in the system. Another issue should be noted is that the first Brillouin zone of silicon is anisotropic. Hence the wave packet propagation direction (group velocity direction $\mathbf{v}_g = \overline{\nabla}_\mathbf{k}(\omega)$) in general is not in the same direction as the wavevector $\mathbf{k}$. For simplicity, we choose a wavevector $\mathbf{k} = [0.1, 0.1, 0] \cdot 2\pi/a$ in a high-symmetry direction to ensure that the wavevector and group velocity are in the same direction. The span of the wave packet ($\sigma$ in Eq. (1) and (2)) is about 20 nm. The incident angle is 45° and the wavelength is 38 Å. Only the acoustic modes are considered, namely TA1, TA2, and LA, which have frequencies of 0.99, 1.42, and 2.26 THz, respectively. From the eigenvectors, we note that TA1 mode is purely in-plane transverse polarization and perpendicular to the wave vector. In comparison, TA2 and LA mode are not purely transverse and longitudinal modes. TA2 mode is mainly out-of-plane but has about 0.5% in-plane square displacement. LA is mainly in-plane but has about 0.5% out-of-plane square displacement. This is due to the fact that the silicon crystal is not symmetric with respect to the (0,0,1) plane (*xy* plane in our simulation). The optical modes are not considered for two reasons. First, they generally do not contribute much to the heat conduction[55]. Second, the wave packet of the optical modes propagates extremely slow, which requires very long simulation time.

## C. Specularity parameter

As mentioned in the introduction, we are quite interested in the energy propagation direction after reflection, because it can tell us how the phonon transport would be affected by the phonon surface scattering. The most widely used model is the specular-diffusive scattering model[14,20], in which the specularity value of *p* means that *p* portion of the incident energy is reflected in the specular direction, with the rest of energy diffusively scattered over all outgoing angle uniformly [20]. In general, the approximate formula for specularity that first derived by Ziman[20] and then extended by Soffer[22] to include angular dependence is used to calculate the specularity:

$$p = \exp(-4\eta^2 k^2 \cos^2 \theta). \quad (4)$$

where $\eta$ is the root mean square (RMS) of surface roughness, *k* is the incident wavevector, and $\theta$ is the incident angle. Here the specularity is only a function of the RMS roughness, wavevector, and the incident angle. Other information, such as wave polarization and the correlation length of the surface are not included in this formula. In fact, Eq. (4) is later been generalized by Ogilvy to include the mode conversion at surface[56]. However, the current form shown in Eq. (4) is more commonly used when considering phonon surface scattering[23–26], probably for simplicity. We noted that for smooth and periodically rough surfaces, the reflected energy is highly focused on certain directions. In such case, a bidirectional reflectance distribution function, which describes how the reflected energy is distributed over the hemisphere upon the reflection surface, is more appropriate to fully describe the angular distribution of reflected energy [38]. For comparison, we still extract the specularity parameter from the portion of energy that is reflected in the specular direction and compare it with Ziman's formula.

Two approaches can be applied to extract the specularity parameter from simulations: from reciprocal space and from real space. In the reciprocal space approach, we can directly project the energy of reflected wave packet into different **k** through Eq. (3) and then we can know the energy at the specular direction (**k** = [0.1, 0.1, 0] · $2\pi/a$). In reality, the peak at this point is so sharp that our reciprocal space resolution is not sufficient to numerically integrate the peak region and accurately obtain the energy. As such, we choose to calculate the specular reflection energy from real space. In this approach, we manually identify the specular reflected wave packet, and choose a region around the wave packet to calculate the total energy. The specularity parameter $p$ in our calculation is just defined as the ratio of the specular wave packet energy to the incident wave packet energy. On the other hand, as usually implied, (1-$p$) should give us the energy that is diffusively reflected. There is also other definition of (1-$p$) based on part of energy that is not coherent reflected[57]. However, technically it is difficult for us to clearly identify the incoherently reflected energy. As such, we assume (1-$p$) is just the faction of energy that is not specularly reflected.

# III. Results and discussions

## A. Phonon scattering at smooth surface

The snapshots of phonon wave packet after reflection for fixed and free boundary are shown in Fig. 2 and Fig. 3, respectively. For phonon scattering at a fixed boundary, different polarizations show different behaviors. The TA1 polarization phonon will split into two major wave packets after scattered at the fixed smooth surface: one propagates in the specular direction and the other propagates at a larger reflection angle of about 77º. The specular one has 52.6% of the incident wave packet energy. For TA2 polarization, the entire wave packet is almost specularly reflected by the fixed surface. For LA, the wave packet mainly splits into two different wave packets with the specular one carrying only 7.6% of the total energy. The other wave packet has an elongated shape. By observation, it propagates slightly backwards at an angle of -10º. For scattering at a free boundary, different behaviors are observed. For TA1, the wave packet is almost specularly reflected at the surface with no other preferred reflected directions. For TA2 modes, the reflected wave packet is mainly in the specular direction, while a small portion of energy is propagating along a particular direction with the angle of about 8º. For LA, the behavior is nearly identical to the case of fixed boundary.

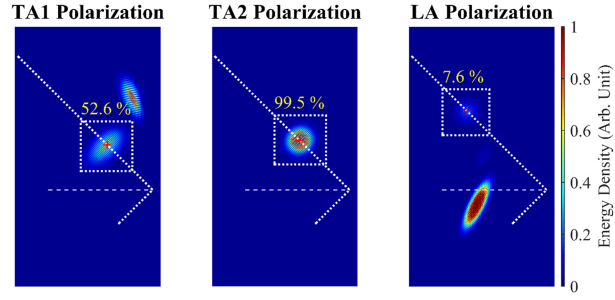

FIG. 2. Energy distribution of TA1, TA2 and LA polarized wave packets after scattered at fixed smooth surface. The cross with red color denotes the calculated center position of the wave packet by assuming the reflection is completely specular.

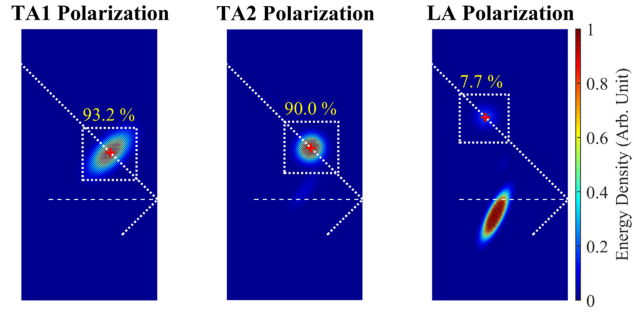

FIG. 3. Energy distribution of TA1, TA2 and LA polarized wave packets after scattered at free smooth surface.

    We further zoom in the wave packets for the fixed boundary case to examine the reflected wave packets in detail, as shown in Fig. 4(a). For the case of TA1, based on the fringes, the wave vector of the wave packet is indeed along the specular direction, and is the same as the propagating direction. From atomic displacements, it is still an in-plane transverse mode (TA1). For the wave packet at larger reflection angle, it is also an in-plane transverse mode, but the propagation direction is not the same as the wavevector direction. As mentioned earlier, the propagation direction of the wave packet is not necessarily the same as the wavevector direction. In order to gain more insights, we also calculated the energy distribution in reciprocal space, as shown in Fig. 4(b). In this figure, the dashed yellow curves show the constant frequency contours from lattice dynamics simulation. As the temperature is 0 K, the scattering should be purely elastic so the wave packet after scattering must fall on these curves. Also, based on the definition of phonon group velocity, its propagation direction must be perpendicular to the constant frequency contours. Note that in order to show everything clearly, we used linear scale for real space plot and log scale for reciprocal space plot. From the left figure of Fig. 4 (b), one can see two peaks in reciprocal space. One is at the wavevector (-0.1, 0.1, 0), which corresponds to the specularly reflected wave packet, and the other is around wavevector (0.02, 0.1, 0), which corresponds to the wave packet at the larger reflected angle. From the constant frequency contours, the group

velocity direction (the gradient direction of constant frequency contours) is also consistent with the results obtained from real space plot. For the TA2 wave packet, it is a single out-of-plane transverse phonon mode in the specular direction (same polarization as TA2), which can be clearly seen from both the real space and reciprocal space plot. The two small peaks in the reciprocal space plot are due to the mode conversion after scattering. Because TA2 is not purely out-of-plane, the small in-plane component allows TA2 mode to be converted into TA1 and LA mode after reflection. Because the component is small, these two wave packets are too weak to be seen in the real space plot. For LA wave, the specular wave packet is still an LA mode, but the other major wave packet becomes a back-propagating transverse wave, corresponding to the reciprocal space peak located at around (-0.25, 0.1, 0). Also, the small out-of-plane component of LA mode allows the reflected wave packets to contain a TA2 mode, which can be seen more clearly in the reciprocal space plot. For each polarization, the reflected wave packets have almost the same frequency as the original incident wave packet, which confirms that the phonon reflection process is completely elastic. Finally, we note that from the reciprocal space plot, the reflected waves all have a $k_y$ = 0.1, indicating that $k_y$ remains unchanged after the scattering process.

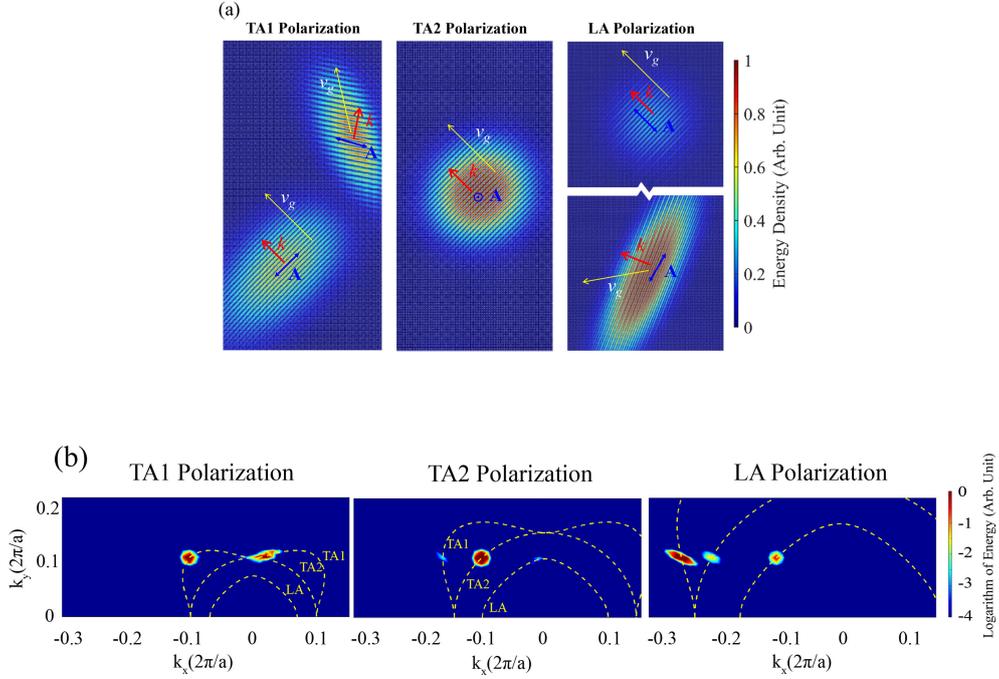

FIG. 4. (a) Enlarged image of wave packets after scattering at fixed smooth surface. Colored arrows represent directions of wave vector (red), group velocity (yellow) and atom displacement (blue). (b) Energy distribution in the reciprocal space for the three different incident polarizations. Both figures are for the fixed boundary case. The dashed yellow curves are the constant frequency contours corresponding to the incident wave packet center frequencies, i.e., 0.99, 1.42, 2.26 THz, respectively. In order to have the best visualization effect, we used linear scale for real space plot and log scale for reciprocal space plot.

Due to the similarity between the long wavelength phonon and elastic wave, it is worthwhile to compare our results with the elastic wave scattering on smooth surface. Silicon is an anisotropic elastic medium so the anisotropic elastic continuum theory should be used [59]. Based on the elastic scattering theory, there could be mode conversion after scattering at a surface, and the boundary condition requires that the wave vector parallel to the surface ($k_y$) must be unchanged after scattering[59]. In fact, this is similar to our results. However, because of the crystal symmetry, the TA2 mode is not purely out-of-plane, and LA mode is not purely in-plane. This makes the mode conversion from TA2 mode to in-plane modes (TA1 and LA) possible. Similarly the LA mode can also convert to TA2 mode. In contrast, for elastic wave, the TA1 and LA wave are purely in-plane, so the mode conversion can only occur between these two modes, while the TA2 mode after reflection should remain a TA2 mode.

## B. Rough surface and specularity parameter

For rough surfaces, we consider two types of surface morphology, namely the sawtooth shape and the square shape. Also, we set surface to be fixed or free. The specularity parameters for sawtooth and square surface at different surface depth $d$ with fixed boundary are shown in Fig 5. For the sawtooth case, the specularity parameters of both TA modes decay as the surface depth increases, similar to the Ziman's model but with quite different values. For LA, the specular component is always quite small. For the square surface, the specularity decays with surface depth, but with more complicated trend. Especially, for the TA2 and LA modes, the specularity at surface depth of 20 Å (close to half of the incident wavelength, which is 19 Å) seems to have some irregular dependence. This is likely due to the interference effect. Since we are using a periodic structure instead of random structure, the periodic surface is similar to diffraction gratings and can induce interference. Some studies also shown that periodic array of pillars at surface of silicon thin film could act as local resonances and alter the phonon spectrum of thin film[60,61]. Our periodic rough surfaces are similar to the surface with array of pillars. Therefore, the local resonant effect could also be the reason for the abnormal dependence of specularity parameter on the depth of surface roughness. For the free surfaces, the specularity results are shown in Fig. 6. The overall trend is decay with increasing surface depth, but the dependence is much more complicated. Again, this could be the combining effects of local resonance at the surface and interference after reflection. Overall, we see that the Ziman's model clearly fails to predict the phonon scattering at these periodic rough surfaces. This model assumes that the medium is continuous and the variation of the roughness at the surface is small compared to the wavelength, which is clearly not applicable for the case when surface roughness and wavelength are similar.

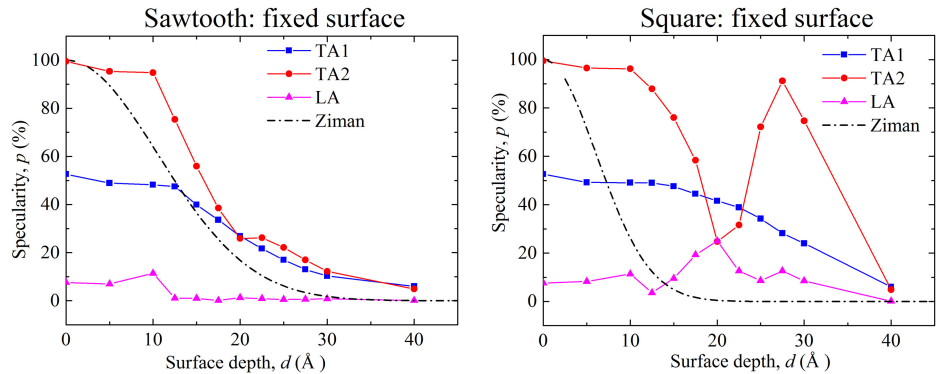

FIG. 5. Specularity parameter for phonon wave packet scattering at fixed surfaces with different surface depth. The dash dot line is based on Ziman's formula.

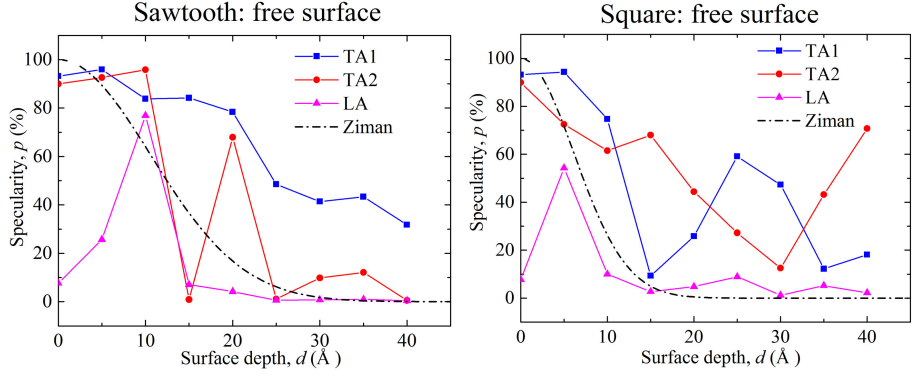

FIG. 6. Specularity parameter for phonon wave packet scattering at free surfaces with different surface depth. The dash dot line is based on Ziman's formula.

## C. Phonon scattering at amorphous surfaces

Next we consider the case where an amorphous layer is attached to the surface. Note that this is quite common in silicon thin films or silicon nanowires, where an amorphous silicon dioxide layer almost always exists on silicon surfaces. In many previous experiments and simulations, the effect of amorphous layer is generally believed to be important to reduce the lattice thermal conductivity of thin films or nanowires[4,62,63]. Here, due to the difficulty in generating a fully relaxed structure for wave packet simulations, we only consider the cases of a smooth interface between silicon and amorphous silicon. Such a simulation may not exactly reproduce the rough silicon surface with amorphous silicon dioxide, but the major physics should be captured.

We first considered the case when the atoms in the amorphous layer are fixed. In this case, the amorphous layer only provides a random perturbation potential field on the surface. We found the results (not shown here for brevity) are essentially similar to the case of scattering on a fixed smooth surface as discussed in Sec III (A). This indicates that the random surface potential itself does not strongly affect the phonon surface scattering, at least for the wavelength we consider here.

We then allowed the atoms in the amorphous layer to move and performed the same simulation. Our simulations indicated that whether the atoms in the outmost layer are fixed or free to move has relatively small effects on the distribution of reflected energy. In Fig. 7(a), we only show the case that the atoms in the outmost layer are free to move. Comparing to the cases of smooth surfaces, no obvious mode conversion is observed when an amorphous layer is attached to the surface. From Fig. 7(a), one can still see a wave packet propagates along the specular direction for all three polarizations. The red cross denotes the calculated position of center of the wave packet at the time instance, assuming the reflection is specular and occurs on the outer surface of the amorphous layer. It can be seen that the red cross almost overlaps with the center of reflected wave packet, indicating that the reflection mainly occurs at the outer surface instead of the interface between silicon and amorphous silicon. However, for the two transverse modes, the specularly scattered wave packet energy is significantly reduced compared to the free boundary case when there is no amorphous layer. Other than the specular wave packet, no other evident wave packets are seen, indicating that the rest of the energy is quite diffusively distributed.

We take the LA wave packet for more detailed analysis. For LA, the specular component is very small, which is similar to the case without amorphous layer. We further calculated the energy distribution in the reciprocal space for LA polarization after surface scattering and the results are shown in Fig. 7(b). The dashed lines in this figure are the contours with frequency equals to the incident LA wave frequency ($f$=2.26 THz), calculated from lattice dynamics simulation. It is clear that the reflected phonons are no longer concentrated on several points, but scattered over all possible directions, while keeping the frequency unchanged.

To understand why the wave packets are more diffusely scattered by the amorphous surface, we further examine the reflection process for the LA polarization, as shown in Fig. 8(a). It can be clearly seen that the wave packet first enters the amorphous layer and then re-emitted into all directions from the amorphous layer. The vibration mode of silicon and amorphous silicon are different[17]. As a result, the well-defined phonon wave packet in silicon may not be supported by the amorphous region. Part of the wave packet energy will convert to random vibration modes in amorphous layer and then slowly emitted out. This process can be further demonstrated by the energy variation in the amorphous layer. As shown in Fig. 8(b), one can see that the energy of the amorphous layer first increases rapidly (the wave packet energy enters the amorphous layer) and then decays gradually (re-emission process).

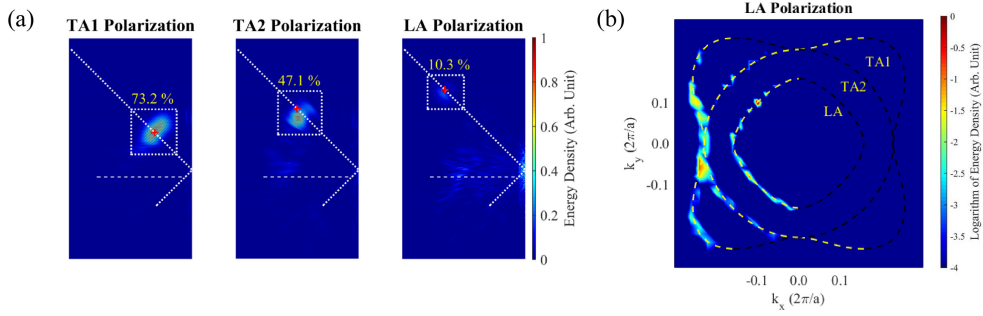

FIG. 7. (a) Energy distribution of the reflected wave at a smooth surface adjacent by an amorphous layer. The red cross denotes the calculated position of center of the wave packet at the time instance, assuming the reflection is specular and occurs on the outer surface of the amorphous layer. (b) The reciprocal space energy distribution for the LA polarization case. The crystalline part of energy at the same time step as Fig 7(a) is used to generate this figure. Dashed lines are constant frequency contours at the same frequency of incident wave packet ($f$=2.26 THz) calculated from lattice dynamics. Based on Eq. (3), the positive or negative sign of wave vector will give the same energy value, therefore the original plot obtained by Eq. (3) is central symmetric. To avoid confusion, we only kept the part that the group velocity has a negative component in the x direction.

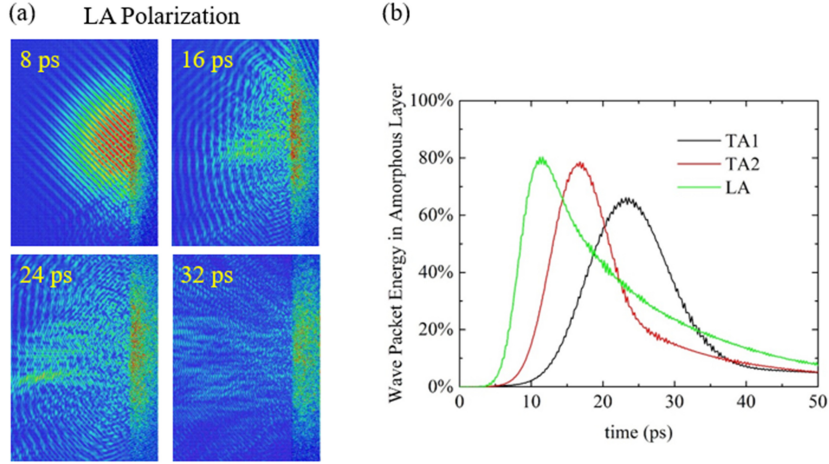

FIG. 8. (a) Snapshots of the scattering process at different instants (b) Energy variation of the amorphous layer.

We also tested the case when the atomic mass of the amorphous layer is doubled to model the case that a silicon surface is coated with amorphous germanium[62]. We note that heavy silicon has quite different vibration spectra compared to silicon[17]. From results shown in Fig. 9(a), we can see that the phonon scattering becomes more diffuse, and it is even difficult for us to extract the specular wave packet, which verify our hypothesis that the more diffuse scattering is due to the vibrational mismatch between silicon and the amorphous layer. The energy variation of the amorphous layer in the heavy amorphous layer is shown in Fig. 9(b), we can see the re-emission process for all the three modes are slower than the amorphous layer with 1 time the mass, likely due to the larger acoustic mismatch between the silicon and amorphous heavy silicon[17].

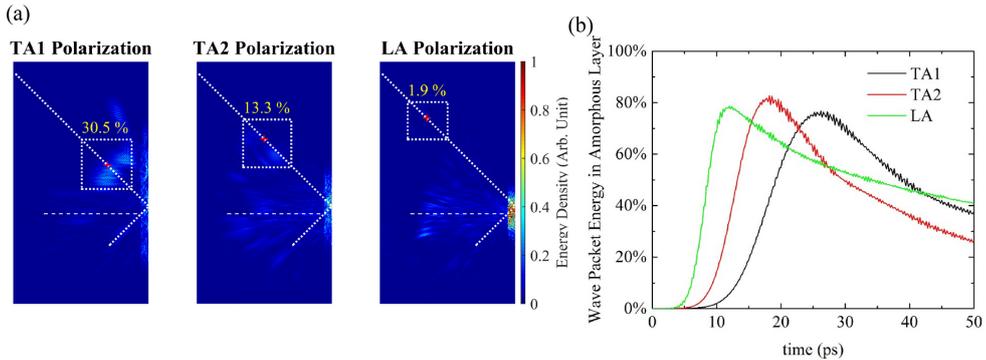

FIG. 9. (a) Energy distribution and specularity parameter at a smooth surface adjacent by an amorphous layer with doubled atomic mass. (b) Energy variation in the amorphous layer.

## IV. Conclusion

In conclusion, we have adopted a two-dimensional wave packet simulation to investigate the phonon surface scattering at different types of surfaces. For smooth surfaces, free and fixed boundary conditions behave differently, and different polarizations also show quite different behaviors. Overall, the out-of-plane modes are nearly specularly reflected, but the in-plane TA and LA mode will split into different phonons after scattering. This behavior is similar to elastic wave reflection on the surfaces of anisotropic materials. For rough surfaces, the specularity parameter is extracted for sawtooth shape and square shape surfaces with different surface roughness. An overall decay in specularity with increasing surface roughness is seen but the trend is complicated and different from the prediction from Ziman's formula. For boundaries with an amorphous layer, there is a specular wave packet after reflection with the rest of the reflected energy propagates quite diffusively. The reason is that the wave packet energy first enters the amorphous layer and then is re-emitted from the amorphous layer. These results allow us to develop a better understanding of the phonon surface scattering process and the thermal transport in nanostructures.

# Acknowledgement

This work is supported by the National Natural Science Foundation of China (No. 51676121). Simulations were performed with computing resources granted by HPC ($\pi$) from Shanghai Jiao Tong University. Fruitful discussion with Prof. Wenjie Wan and Prof. Yanfeng Shen are gratefully appreciated. C. Shao would like to thank Prof. I-Ling Chang and Dr. Wee-Liat Ong for proofreading the manuscript.